\documentclass[aps,prl,twocolumn,superscriptaddress,longbibliography,showkeys,floatfix]{revtex4-1}
\usepackage{pifont}
\usepackage{color}
\usepackage{amsmath}
\usepackage{graphicx}
\usepackage{lineno}

\usepackage{type1cm}
\usepackage{eso-pic}

\def\tt{TT}
\def\het{$\text{THe}^+$}
\def\heh{$\text{HHe}^+$}
\def\hplus{$\text{H}^+$}
\def\tplus{$\text{T}^+$}

\def\hminus{$\text{H}^-$}
\def\heplus{$\text{He}^+$}
\def\heplusplus{$\text{He}^{++}$}
\def\hh{HH}

\begin{document}

%\preprint{}

\title{Beta decay of molecular tritium}

\author{Y.-T.~Lin}\email[]{tinglin194@gmail.com}
\affiliation{Center for Experimental Nuclear Physics and Astrophysics and Department of Physics, University of Washington, Seattle, WA 98195}

\author{T.~H.~Burritt}
\affiliation{Center for Experimental Nuclear Physics and Astrophysics and Department of Physics, University of Washington, Seattle, WA 98195}

\author{C.~Claessens}
\affiliation{Institute of Physics, Johannes Gutenberg University Mainz, 55099 Mainz, Germany} %Staudinger Weg 7,

\author{G.~Holman}
\affiliation{Center for Experimental Nuclear Physics and Astrophysics and Department of Physics, University of Washington, Seattle, WA 98195}

\author{M.~Kallander}
\affiliation{Center for Experimental Nuclear Physics and Astrophysics and Department of Physics, University of Washington, Seattle, WA 98195}

\author{E.~Machado}
\affiliation{Center for Experimental Nuclear Physics and Astrophysics and Department of Physics, University of Washington, Seattle, WA 98195}

\author{L.~I.~Minter}
\affiliation{Center for Experimental Nuclear Physics and Astrophysics and Department of Physics, University of Washington, Seattle, WA 98195}

\author{R.~Ostertag}
\affiliation{Institute of Experimental Particle Physics (ETP), Karlsruhe Institute of Technology (KIT),  76131 Karlsruhe, Germany} %Wolfgang-Gaede-Str. 1,

\author{D.~S.~Parno}\email[]{dparno@cmu.edu}
\affiliation{Carnegie Mellon University, Pittsburgh, PA 15213}

\author{J.~Pedersen}
\affiliation{Center for Experimental Nuclear Physics and Astrophysics and Department of Physics, University of Washington, Seattle, WA 98195}

\author{D.~A.~Peterson}
\affiliation{Center for Experimental Nuclear Physics and Astrophysics and Department of Physics, University of Washington, Seattle, WA 98195}

\author{R.~G.~H.~Robertson}\email[]{rghr@uw.edu}
\affiliation{Center for Experimental Nuclear Physics and Astrophysics and Department of Physics, University of Washington, Seattle, WA 98195}

\author{E.~B.~Smith}
\affiliation{Center for Experimental Nuclear Physics and Astrophysics and Department of Physics, University of Washington, Seattle, WA 98195}

\author{T.~D.~Van~Wechel}
\affiliation{Center for Experimental Nuclear Physics and Astrophysics and Department of Physics, University of Washington, Seattle, WA 98195}

\author{A.~P.~Vizcaya~Hern\'{a}ndez}
\affiliation{Carnegie Mellon University, Pittsburgh, PA 15213}

\collaboration{The TRIMS Collaboration}
\noaffiliation

\date{\today}

%\linenumbers

\begin{abstract}
The beta decay of tritium in the form of molecular TT  is the basis of sensitive experiments to measure neutrino mass.  The final-state electronic, vibrational, and rotational excitations modify the beta spectrum significantly, and are obtained from theory. We report measurements of the branching ratios to specific ionization states for the isotopolog HT.  Two earlier, concordant measurements gave branching ratios of HT to the bound \heh\ ion of 89.5\% and 93.2\%, in sharp disagreement with the theoretical prediction of 55-57\%, raising concerns about the theory's reliability in neutrino mass experiments.  Our result, 56.5(6)\%, is compatible with the theoretical expectation and disagrees strongly with the previous measurements. 

\end{abstract}

\keywords{Beta decay, neutrino mass, tritium, molecular structure theory}

\maketitle
%\tableofcontents

%%\section{Introduction}c

The discovery of neutrino oscillations and mass \cite{Fukuda:1998mi,Ahmad:2001an} signals a contradiction to a prediction of the minimal standard model, and  opens a window to the physics that lies beyond.  Oscillations link the squares of the masses via their differences, but do not give values for the masses themselves.  The recent KATRIN result from tritium beta decay gives an upper limit of 1.1 eV on each mass \cite{PhysRevLett.123.221802}.  An intensive effort continues to determine the masses experimentally \cite{drexlin:2013}.   A laboratory measurement will constrain rates for the hypothesized neutrinoless double beta decay process and help disentangle correlated parameters in cosmological models \cite{Choudhury_2018}.

The most sensitive direct method for probing the neutrino mass is by examining the beta spectrum of  tritium near the end point.   The shape of the beta spectrum there is affected by molecular excitations, which must  be calculated with great precision in order to be confident in the additional contribution of a non-zero neutrino mass \cite{Bodine:2015sma}.  While these `final-state' calculations can in principle be taken to arbitrary accuracy since the force is known, in practice approximations are necessary. Indeed, the first precision experiments with TT \cite{robertson:1991aa,stoeffl:1995aa} produced apparently negative values of $m_\nu^2$, a result that has been traced \cite{Bodine:2015sma} to inadequacies in the final-state theory in use at the time (for brevity we write H for $^1$H, T for $^3$H, and He for $^3$He).  Those shortcomings have been eliminated in the work of Saenz, Jonsell, and Froehlich \cite{jon99,saenz00} using overlap integrals and a geminal basis for the parent and daughter molecules.  Even so, the root-mean-square spread of final states in the region of interest must be known to 1\% in order for KATRIN \cite{PhysRevLett.123.221802} to meet its 0.2-eV sensitivity goal.  

One puzzling discrepancy remains.  In the 1950s, two experimental studies of the molecular ions made in the beta decay of HT and \tt{} were carried out using mass spectrometers \cite{snell57,wex58,wexler:1969}, and both indicated that 90 to 95\% of decays lead to the bound molecular ion \heh\  or \het.  Theory, however, predicts the fraction to be at most 57\% (Table~\ref{tab:intro}).    The electronic ground state of \heh{} and \het{} is bound by 2 eV, but lepton recoil can excite rotational-vibrational states that are energetically unbound.   Most of the latter states are hindered from dissociation by their angular momentum, and thus quasibound. 
\begin{table}[hbt]
\centering
\caption{ Branching ratio to the bound molecular ion for HT and \tt.}
\label{tab:intro}

\begin{tabular}{cccccc}
\hline \hline
   	 &  Snell {\em et al.} & Wexler & \multicolumn{3}{c}{\phantom{a}Theory \phantom{a}}\\   
 & \cite{snell57} & \cite{wex58}& \multicolumn{3}{c}{\cite{jon99,saenz00}} \\
Molecule & & & Quasibound & Bound & Total \\
\hline 
HT  & 0.932(19)& 0.895(11) & 0.02 & 0.55 & 0.57 \\ 
\tt  & -- & 0.945(6) & 0.18 & 0.39 & 0.57 \\
\hline \hline
\end{tabular}
\end{table}
% \begin{table}[hbt]
% \centering
% \caption{ Branching ratio to the bound molecular ion for HT and \tt.}
% \label{tab:intro}

% \begin{tabular}{ccccc}
% \hline \hline
%  Molecule  	 & \phantom{a} Snell {\em et al.}\phantom{a} &\phantom{a} Wexler\phantom{a} & \multicolumn{2}{c}{\phantom{a}Theory \phantom{a}}\\   
%  & (Ref.~\cite{snell57}) & (Ref.~\cite{wex58})& \multicolumn{2}{c}{(Ref.~\cite{jon99,saenz00})} \\
%  & & & Quasibound & Bound \\
% \hline 
% HT  & 0.932(19)& 0.895(11) & 0.02 & 0.55 \\ 
% \tt  & -- & 0.945(6) & 0.18 & 0.39 \\
% \hline \hline
% \end{tabular}
% \end{table}

The theory might indeed be in error, one of many hypotheses advanced  \cite{Bodine:2015sma} to explain the disagreement.   Strictly, the theory applies to the endpoint whereas the experiments integrate over the entire beta spectrum. Unidentified systematic errors may have affected the measurements.    However, none of these explanations seems likely to accommodate such a large disagreement.  

We present new measurements of the branching ratios for HT and TT with a novel time-of-flight mass spectrometer, TRIMS (Tritium Recoil Ion Mass Spectrometer).  Modern neutrino mass experiments rely on TT, but the underlying theory for HT and TT decay is the same,  predicting a fraction of 0.57 (the distinction between bound and quasibound is immaterial for neutrino mass experiments).  The heteronuclear parent HT permits clear separation of all ionic final states, and the quasibound fraction is smaller, making HT the better choice for a  decisive experimental test of the theory.

The beta decay of tritium in the form of HT gas produces the positively charged ions \hplus, \heplus, \heplusplus, and \heh.  The negative ion  \hminus\  can also be produced but is expected to be rare.  When \tt{} is also present, \tplus and \het{} are produced. Electrons produced include the beta itself and 0, 1, or 2 shakeoff electrons.  

\begin{figure}[t]
	\centerline {
	\includegraphics[width=3.375in]{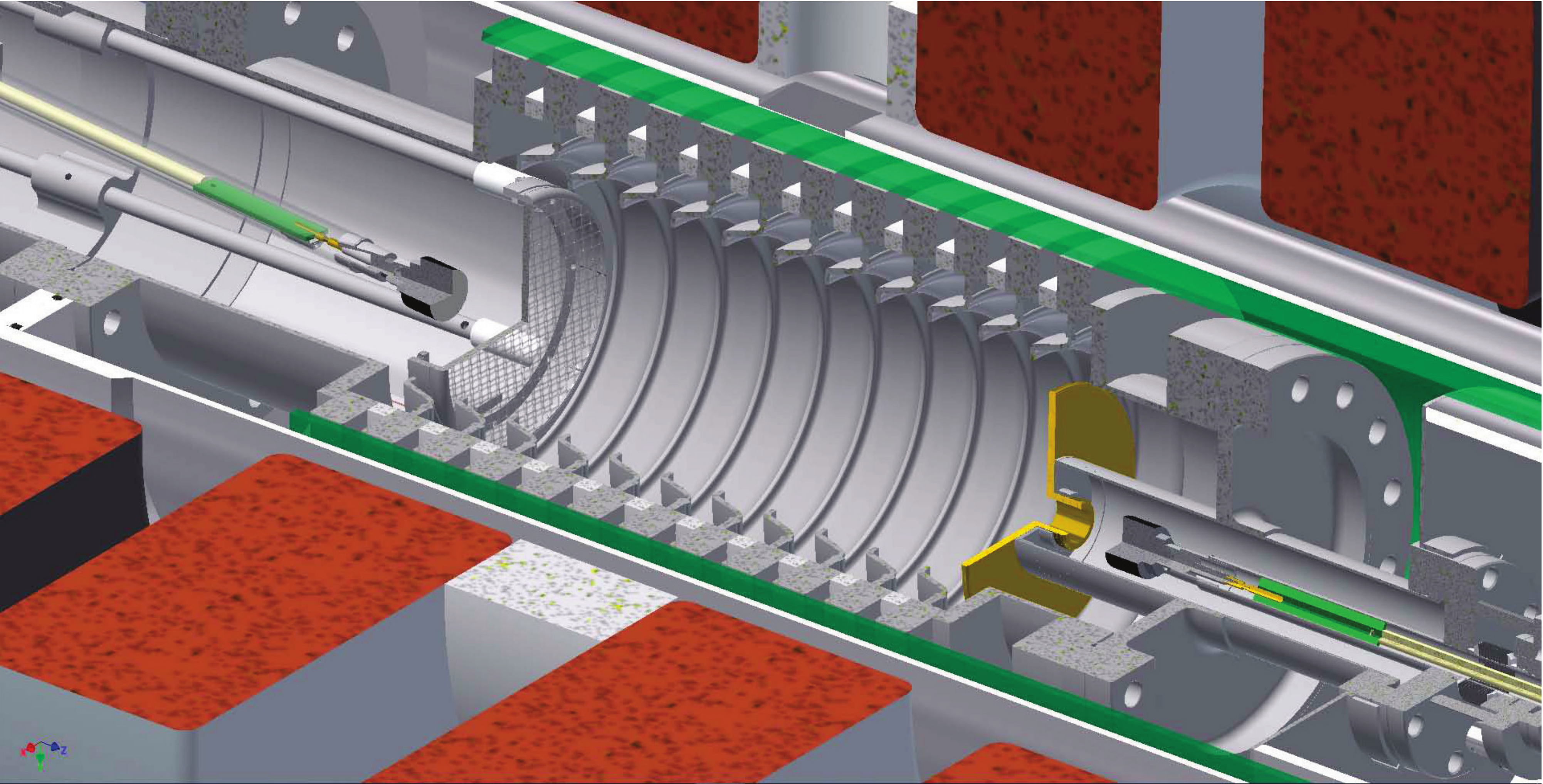}}
	\caption[]{Cutaway view of the decay chamber inside the insulating silica tube, aluminum tube, and  magnet coils. The 11-gap acceleration structure is 234 mm long. The beta detector is  on the right behind the gold-plated high-voltage electrode; the ion detector is to the left behind the biased mesh.}  
	\label{fig:cutaway}
\end{figure}

  In TRIMS, ions and electrons move under the influence of collinear, uniform magnetic and electric fields toward silicon detectors located at either end of a decay chamber (Fig.~\ref{fig:cutaway}).  The start time is set by the arrival of an electron in the `beta' detector at the anode end, and the position in the chamber where the decay occurred is determined by the energy $K_{\rm ion}$ acquired by a positive ion en route to the `ion' detector at the cathode end.  The ion mass-to-charge-squared ratio is then given by 
\begin{eqnarray}
\frac{m}{q^2}&=& t^2\frac{E^2}{2K_{\rm ion}}, \label{eq:mass}
\end{eqnarray}
where  $t$  is the ion's flight time and $E$ is the uniform electric field in the chamber.  Not shown, but  included in the analysis, is a correction for a drift space in front of the ion detector. 

A Monte Carlo simulation using the GEANT4 toolkit~\cite{Agostinelli2003,Allison2016} was developed to guide design and study the influence of various systematics.  The simulated energy and timing at the detectors are  smeared  by  measured detector resolutions. Also incorporated in the simulation are electron backscattering \cite{DONDERO201818} and the deposited ion energy predicted by SRIM \cite{Ziegler2010}.

The apparatus \cite{BodinePhD} at the  University of Washington consists of a metal-sealed ultra-high-vacuum system with an ion pump, two SAES  getter pumps with ST-101 alloy, a liquid nitrogen cold trap, and a turbomolecular pump.   After baking, it reaches a base pressure of  $\sim10^{-9}$ mbar.   

 The decay chamber (Fig.~\ref{fig:cutaway}) is a National Electrostatics acceleration column, made of alternating Kovar alloy and alumina rings brazed together. Shaped aluminum rings mounted to the Kovar rings and biased by a chain of 76 1-G$\Omega$ 1\% resistors establish a uniform voltage gradient of 281 V/mm. Four external magnet coils produce a 0.236-T magnetic field uniform to $\pm0.5$\% throughout the decay chamber.  COMSOL calculations \cite{COMSOL} showed that the field was not significantly affected by the Kovar rings.

At the anode end is a stainless-steel disk, gold-plated  to reduce  emission of secondary ions, with a circular opening for the beta detector to look through. At the cathode end is an 85\%-transmitting  stainless-steel mesh mounted under tension on a metal ring.  It is electrically isolated and held 100~V below ground to suppress secondary electrons from the ion detector, located 29 mm behind the mesh.  

With the chamber isolated, commercially supplied gaseous tritium is introduced through a leak valve to a partial pressure of $\sim3\times10^{-8}$ mbar in normal operation.  The total pressure, mostly H$_2$, is monitored via a spinning-rotor gauge to avoid pumping of the gas by ionization gauges, and rises $\sim10^{-5}$ mbar/hr. Slow-control data from devices are sent to a PostgreSQL database on a remote server.

Silicon detectors with 50-mm$^2$ area and 0.5-mm thickness (Canberra PD50-11-500AM), are mounted on movable re-entrant arms containing custom, low-noise, miniature preamplifiers mounted directly on the feedthrough pin for minimal connection capacitance, and cooled to $\sim 15^{\circ}$C with CO$_2$ Joule-Thomson coolers. By means of two translation stages, the ion detector can be moved up to 25 mm off-axis in any direction to measure radial distributions.  The detector signals, transmitted through fiber optics, are digitized with a 12-bit 250-MHz digitizer that is triggered when either signal is above a discriminator threshold.  The digitizer outputs are read out with the ORCA data acquisition software~\cite{how04} and translated into ROOT~\cite{Brun:ROOT97} format. To obtain the signal amplitudes, two consecutive trapezoidal filters are applied on each   waveform~\cite{jor94,Amsbaugh2015}. The trigger time for each waveform is obtained by fitting a Woods-Saxon (or logistic) function \cite{1954PhRv...95..577W} to the waveform. The fit reduces the effect of electronic noise at low pulse amplitudes.  The large-signal (90 keV) timing resolution is 4~ns FWHM, which is broadened by noise to 35~ns for signals at 10~keV.     

Calibrated with $^{241}$Am gammas, the beta and ion detector energy resolutions were 2.46 and 2.05 keV FWHM, respectively.    The ionization-yield energy calibration of the detectors is established with gammas, accelerated secondary electrons, and electronic pulser signals.   Unlike gammas, ions lose energy in semiconductor detectors in the dead layer and to nuclear scattering.  To account for these effects, the stopping power tables of SRIM \cite{Ziegler2010} are numerically integrated to derive the ionization yield, or `detected energy', as a function of incident energy for each particle type.  The yields are fitted with cubic polynomials to give the detected energy for a given true incident energy over the range of interest.  The converse gives the corrected, true energy $K_{\rm ion}$ for each ion species, with fitting errors $\lesssim 0.1$ keV.   The ion detector gain was monitored {\em in situ} from two-ion H$^+$+He$^+$ events wherein the initial beta is not detected and the \heplus{} strikes the mesh, dislodging a secondary electron that is detected.  The protons passing through to the detector form a continuous energy band terminating at true energies of 59.7 keV, the acceleration potential across the chamber.  That constraint also yields the dead layer thickness, 75(20) nm.

%\section{Calibration, simulation and analysis tools}

\begin{figure}[ht]
	\centerline {
	\includegraphics[width=3.375in]{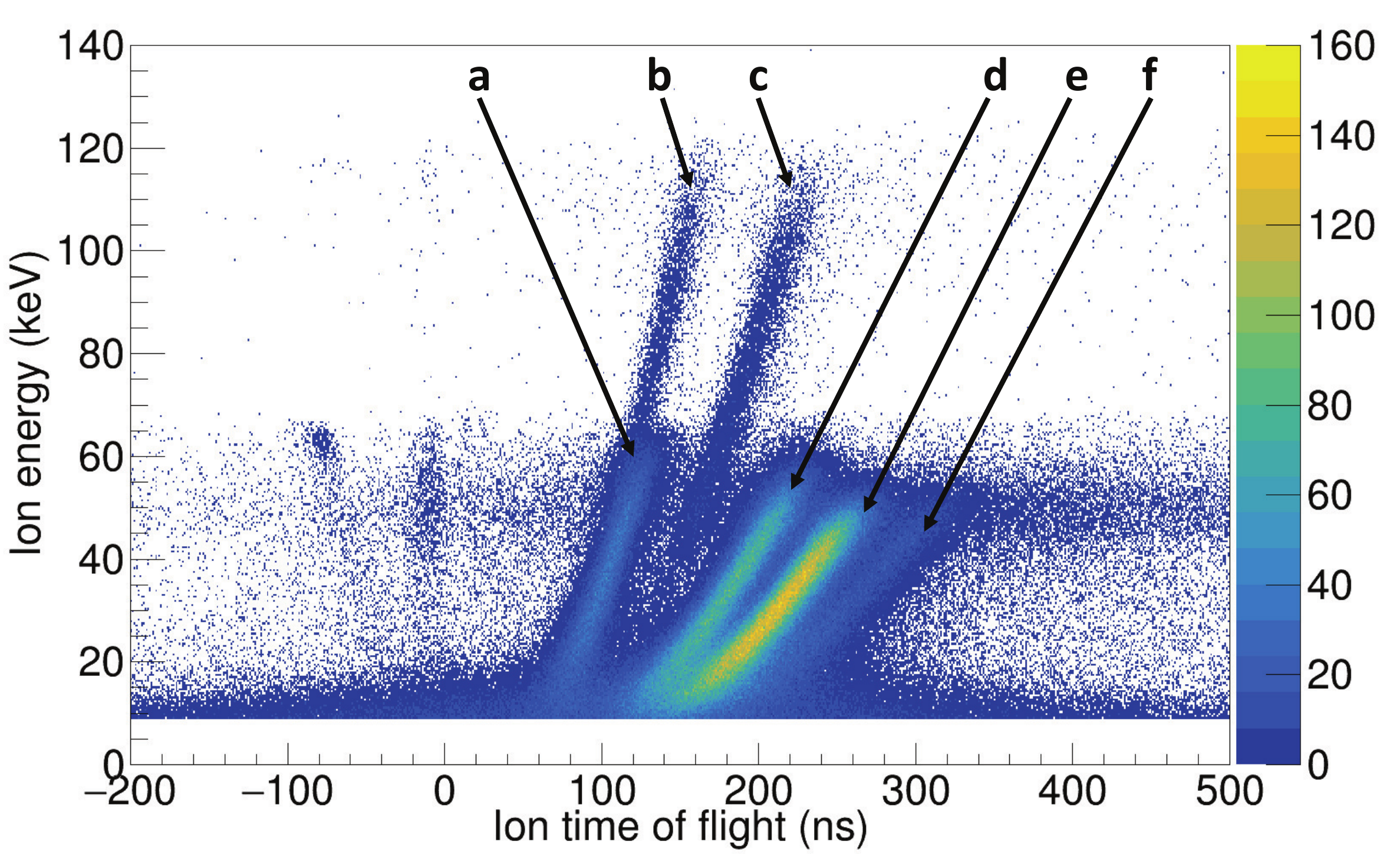}
	}
	\centerline {
	\includegraphics[width=3.375in]{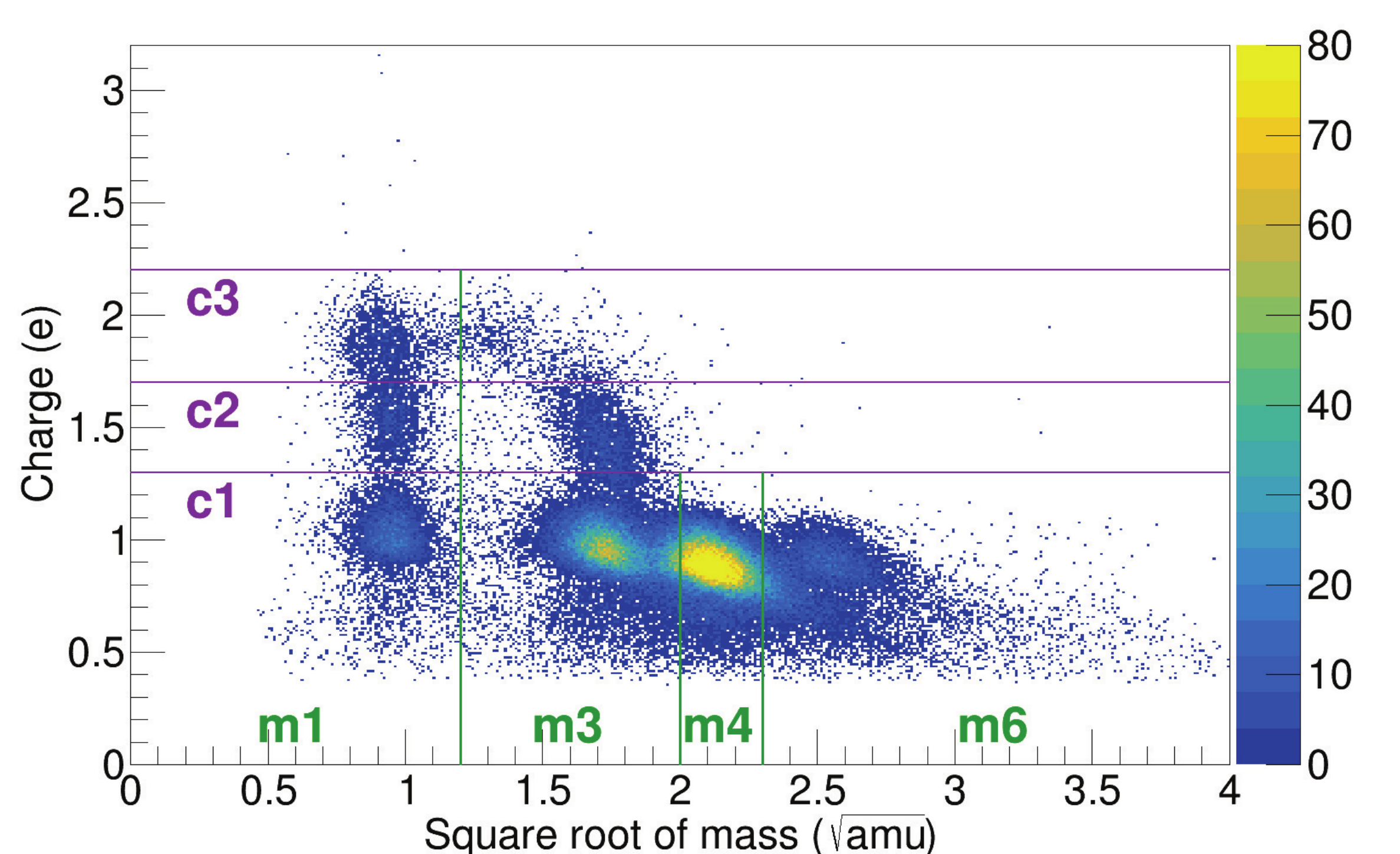}
	}
	%\centerline {
	\includegraphics[width=3.375in]{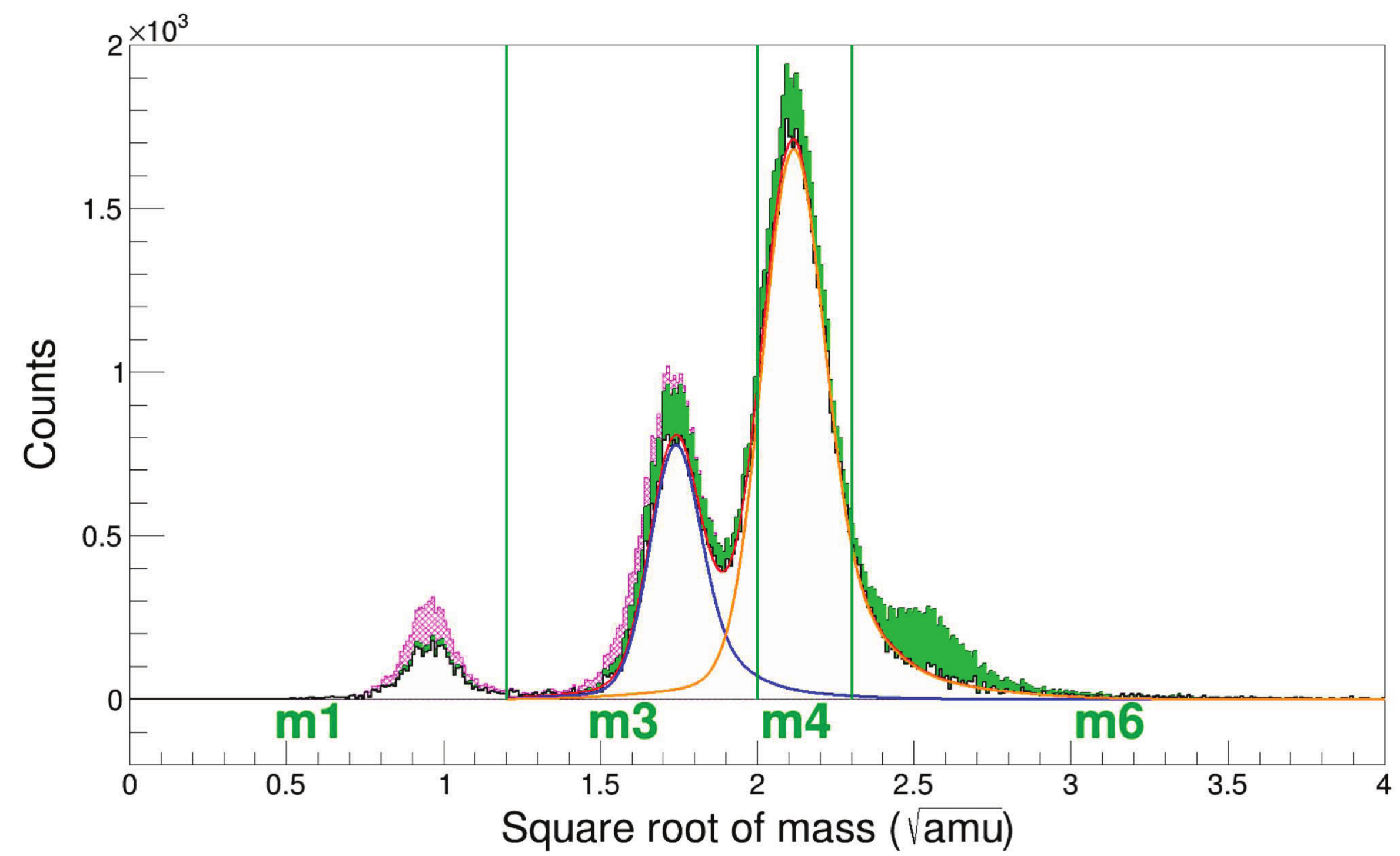}
	%}
	\caption[]{Raw data from HT-rich source.  Top: a), d), e), f) -- the main ion bands associated with masses 1, 3,  4, and 6; b) -- doubly-charged $\text{He}^{++}$ band; c)  two-ion band with $\text{He}^{+}$, $\text{T}^{+}$, and $\text{H}^{+}$. Bands to the left of time zero are secondary emission bands.   Bin size: (1 ns, 0.4 keV).
	Center: Derived values of charge and mass for each event. Bin size: (0.01, 0.01). Bottom: Mass spectrum of the main  charge-1 bands after removing contributions from \tt{} (solid green) and from two-ion detection branches 6e and 6f (hatched magenta).  The mass-3 and 4 peaks are each fit with three Gaussians; the results are shown. Bin size: 0.01.
	}
	\label{fig:data}
\end{figure}

For the purposes of calculating charge and mass spectra (Fig.~\ref{fig:data}) $K_{\rm ion}$ for He was used.   Once events had been grouped by charge and mass, the  effective fiducial `volume' (FV) for each species was  determined from the range of true energies accepted from the acceleration chamber, which depends (slightly) on the particular ion because the energy cut, 20 to 40 keV, is placed on detected, rather than true, energy.    Ions can be backscattered and lost, effects also modeled with the aid of SRIM.  The loss percentages $b_i$ and fiducial volumes are included in Table~\ref{tab:branch}.

During a typical 1-hr HT data run, \tt{} was introduced to the desired count rate, limited to $\sim 200$ s$^{-1}$ to avoid  deadtime and pileup.  HT is created by exchange between the introduced \tt{} and the \hh{} outgassing from the walls, catalyzed by platinum-group metals present in vacuum-gauge filaments~\cite{NortonRichards-HDCatalysis-1974}.  The gauges could be valved off from the rest of the system, allowing measurements to be taken with predominantly HT or predominantly \tt{}.  Catalytic conversion proceeded with a time constant of 8 min to an asymptotic HT activity fraction of 95\%.  

The decay channels for HT are listed in Table~\ref{tab:branch}. Most decays lead to the single-charge `main' branches, 2, 3, and 4.  In cases where two positive ions and two or more electrons are produced in the final state, the different detectable configurations  are listed under the primary channel responsible for producing them (e.g., 6a, 6b, etc.).  These combinatorial situations are treated in analysis. Only charged-particle final states are included: for the bound-state beta decay branch, 1,  Bahcall \cite{PhysRev.124.495} calculates 0.69\% in the case of a bare tritium ion.

Fundamentally, the branching ratio is the fraction of coincidence events with a beta  that lead to a specific final channel  $i$ (see Table~\ref{tab:branch}).  Details of the multistep extraction of branching ratios are given in  \cite{LinPhD}. 

\begin{table}[htb]
   \centering
   %\topcaption{Table captions are better up top} % requires the topcapt package
   \caption{Decay and detection channels of HT.  Also shown are ion backscatter loss percentages $b_i$ and the fiducial `volume' FV in keV for each channel.}
    \label{tab:branch}
   \begin{tabular}{@{} llrr @{}} % Column formatting, @{} suppresses leading/trailing space
    \hline             \hline
$i$ &Channel&\phantom{aaaa}$b_i$ (\%)&FV (keV)\\
      \hline
\multicolumn{4}{l}{Zero electron (bound-state beta decay)} \\
1. \phantom{a} &  ${\rm He}+{\rm H} $ &\\
      \hline
 \multicolumn{4}{l}{One electron} \\
2. & ${\rm HeH}^++e^- $  &3.2&20.74\\
3. & ${\rm He}^++{\rm H}+e^- $&1.1&20.11\\
4. &${\rm He}+{\rm H}^++e^- $ &0.3&18.12\\
5. & ${\rm He}^{++}+{\rm H}^{-}+e^-$&1.1, 0.3&20.11/2\\
      \hline
 \multicolumn{4}{l}{Two electrons (one shakeoff)} \\
6.  & ${\rm He}^++{\rm H}^++2e^- $ &&\\
\phantom{a}a&${\rm He}^++{\rm H}^++2e^- $ &1.1, 0.3&20.23/2\\
\phantom{a}b& ${\rm He}^++{\rm H}^++1e^- $ &1.1, 0.3&20.23/2\\
\phantom{a}c& ${\rm He}^++2e^- $ &1.1&20.11\\
\phantom{a}d& ${\rm H}^++2e^- $ &0.3&18.12\\
\phantom{a}e& ${\rm He}^++1e^- $&1.1&20.11\\
\phantom{a}f& ${\rm H}^++1e^- $ &0.3&18.12\\
7. &${\rm He}^{++}+{\rm H}+2e^- $ &&\\
\phantom{a}a& ${\rm He}^{++}+{\rm H}+2e^- $ &1.1&20.11/2\\
\phantom{a}b&${\rm He}^{++}+{\rm H}+1e^- $ &1.1&20.11/2\\
      \hline
\multicolumn{4}{l}{Three electrons (two shakeoff)} \\
8. &${\rm He}^{++}+{\rm H}^++3e^- $ &&\\
\phantom{a}a&${\rm He}^{++}+{\rm H}^++3e^- $ &1.7, 1.5&20.11/3\\
\multicolumn{4}{l}{\phantom{a} +11 combinations} \\
\hline \hline
\end{tabular}
\end{table}

The most significant correction arises from initial ion momentum, which causes the radial distribution in the ion-detector plane to depend on the specific decay channel. These distributions are not known {\em a priori}: the coincidence rates are therefore measured by scanning the ion detector radially from the axis.  The measured distributions are termed Raw Count Functions (RCF). Because the ion detector is of non-negligible size, a point-spread function (PSF) is calculated from the geometry in order to deconvolve the underlying radial distributions.  The integrals of those distributions are the corrected quantities needed to form the branching ratios.  Other corrections are noted below.

%\section{First results with HT}
Coincidence data from TRIMS are shown in Fig.~\ref{fig:data}.  Bands produced by the decay channels shown in Table~\ref{tab:branch}, particularly from the singly charged ions with mass 1, 3 and 4~amu, may be seen.  In addition, minor bands from charge-2 decays marked $b$ and $c$, a mass-6 band from residual \tt, marked $f$, and secondary-emission bands to the left of time zero can be seen.   The diffuse horizontal bands with ion energies $\sim 50$ keV or $<10$ keV are from decays that occur in the non-accelerating regions near the detectors. Such events are filtered out via the 20-40 keV ion-energy FV cut.   Some runs were affected by intermittent data-acquisition problems. Affected periods were removed in analysis. 

 Charge-mass spectra (Fig.~\ref{fig:data}, center) were then generated with Eqs.~\ref{eq:charge} and~\ref{eq:mass}.  The effective charge is:
\begin{eqnarray}
q_{\rm eff} &=& ( K_{\rm ion} + K_\beta-K_\beta^0)V^{-1}, \label{eq:charge}
\end{eqnarray}
where $K_\beta$ is the detected electron energy and $V$ the acceleration voltage (59.7 kV).  The unknown initial beta kinetic energy $K_\beta^0$  broadens the charge distribution, but  separation into charge groups is still possible.  For sorting events, $K_\beta^0$ is fixed at 3 keV.   

The plane is subdivided into cells within which the events are predominantly from single decay channels. Cross-contamination between neighboring cells is corrected on the charge axis according to the Geant4 simulation and on the mass axis by the Gaussian fits to the main peaks.  

In the charge-mass plane one sees groups with $q_{\rm eff}=1$ and 2, but also 1.5.  The latter are from two-electron branches 6 and 7 where one ion or one electron is missed. For branch 6 with two ions and two electrons the branching ratio is quantified by treating one of the ions (either one) as though it were a neutral spectator, like the main branch examples 3 and 4. For example, the He$^+$ is detected either by itself or in the company of the spectator H$^+$, and the branch numerator is the sum of those detection channels.  Similarly, if both electrons are detected, we are assured that one is a beta.  When only one is detected, either because the other backscattered and was lost, or  because near the detector edge, one or the other missed the detector, only half of those events have a valid beta.  The electron-loss corrections are determined by simulation.  When one ion and one electron are both lost, the event becomes indistinguishable experimentally from a main-band event.  In this case, the corrections, which are scan-position dependent, are calculated from one-ion and two-ion data where both electrons are detected, coupled with electron-loss simulations. 

We take data with different mixtures of HT and TT.  By subtraction of the appropriately normalized contribution of one mixture from the other mixture, pure HT and pure TT spectra are obtained.   The relative normalization is determined from unique spectral features, namely the two-ion branches \hplus{}+\heplus{} and \tplus{}+\heplus{} from HT and TT, respectively, that produce a secondary electron at the mesh from the impact of one ion.  Events of this kind form unique, completely isolated groups on the time-of-flight axis that do not require analysis beyond the numbers of events in each group.   Pure HT or TT spectra result when the normalization completely removes two-ion events unique to the other isotopolog.   The elimination of the mass-6 peak in the bottom spectrum of Fig.~\ref{fig:data} gives  confirmation.

Events with only the secondary electron detected yield ``beta''-ion times that are small or even negative (Fig.~\ref{fig:data}, top).  These events are not to be included in the branching ratios because they lack a beta.

The  branching ratios obtained are listed in Table~\ref{tab:branchratios}.  The upper limits for branches 5 and 8 represent the numbers of events in the appropriate charge-mass cells, here deemed to be background rather than signal.  Statistical uncertainties are listed in column 3.  RCF uncertainties are additional aggregated uncertainties  prior to deconvolution, and include contributions from  random coincidences, \tt{} spectrum subtraction,  the cross-contamination of neighboring cells by charge and mass peaks, modeling of  electron backscattering and losses, and manual scan-position setting. Point-spread function uncertainties encompass the dimensionality, scan step size, and zero offsets of the deconvolution matrix.  The FV for each branch (Table~\ref{tab:branch}) is subject to energy calibration uncertainties. For  charge-2 branches FV is half as large as for charge-1 branches (indicated by ``/2'' in Table~\ref{tab:branch}).
\begin{table}[ht]
   \centering
   \caption{Branching ratios and uncertainties for decay channels of HT.}
    \label{tab:branchratios}
   \begin{tabular}{@{} llrcccccc @{}} 
\hline \hline
$i$ &Channel&\multicolumn{6}{c}{Uncertainties (absolute \%)} &Branch\\
&&Stat.&RCF&PSF&FV&DT&Total&(\%)\\
\hline
 \multicolumn{4}{l}{One electron} \\
2. & ${\rm HeH}^+ $  &0.1&0.39&0.14&0.35&0.08&0.55&56.51(55)\\
3. & ${\rm He}^++{\rm H} $&0.1&0.27&0.05&0.28&0.05&0.41&24.98(41)\\
4. &${\rm He}+{\rm H}^+$ &0.09&0.4&0.17&0.09&0.01&0.45&5.64(45)\\
5. & ${\rm He}^{++}+{\rm H}^{-}$&&&&&&&$<0.021$\\
\hline
 \multicolumn{4}{l}{Two electrons} \\
6.  & ${\rm He}^++{\rm H}^+ $ &&&&&&&\\
&from  ${\rm He}^+ $ &0.19&0.41&0.09&0.15&0.02&0.49&11.01(49)\\
&from  ${\rm H}^+ $ &0.17&0.37&0.09&0.13&0.02&0.44&10.43(44)\\
7. &${\rm He}^{++}+{\rm H} $ &0.12&0.07&0.16&0.05&0&0.21&2.16(21)\\
\hline
\multicolumn{4}{l}{Three electrons} \\
8. &${\rm He}^{++}+{\rm H}^+$ &&&&&&&$<0.045$\\
\hline \hline

   \end{tabular}
  
   \end{table}

Misalignment of the electric and magnetic fields would cause a departure from rotational symmetry, but was measured to be negligibly small, 4.7(7) mrad. Ion backscattering corrections and small corrections for the dependence on scan position of the scan step size, magnetic field, dead-layer loss, and detector acceptance  were applied and did not contribute significant uncertainties.  A search for $^2$H$^+$ ions yielded 0.3(2)\% of H$^+$, and representative uncertainties are listed under DT in Table~\ref{tab:branchratios}.  Loss of ions to charge exchange was estimated to be less than 0.25\%. 

For TT decay, the measured branch to mass-6 ions is 50.3(15)\%, which, like HT, disagrees strongly with the measurement of Wexler \cite{wex58}, and agrees with the theoretical range of 39 to 57\% (Table~\ref{tab:intro}).  We see clear evidence for dissociation in flight of mass-6 ions from quasibound states as will be described in a forthcoming paper, but theory does not predict their lifetimes, only the range.

%\section{Discussion and conclusion}

In conclusion, the mass spectroscopic measurements of \cite{snell57,wex58} have for 60 years been the only data on the branching ratio of HT and \tt\ to bound and unbound molecular ions.  The profound discrepancy of those experiments with theory would imply either a dramatic and unexpected failure of the sudden approximation \cite{migdal1941ionization}  at relatively high beta energies, or some hindrance of dissociation by 5 orders of magnitude so that radiative decays from highly excited transient molecular states could dominate.

We have measured the branching ratios over the entire beta spectrum, as did Snell {\em et al.}~\cite{snell57} and Wexler \cite{wex58}, and find strong disagreement with the results of both experiments.  In contrast, our results are in accord with theory even over this full range of beta energy.  The source of the disagreement is not known.  It may be due to the momentum acceptance of the instruments as Wexler himself suggested \cite{wex58}.

With the present results, the last known disagreement between experiment and the theory of the final-state distribution in tritium beta decay is removed, providing support for the theoretical analysis of neutrino mass experiments such as KATRIN \cite{PhysRevLett.123.221802} and Project 8 \cite{Asner:2014cwa} that make use of molecular tritium.

%\end{document}
\begin{acknowledgments}
The authors wish to thank Alejandro Saenz and Sanshiro Enomoto for valuable discussions.  This material is based upon work supported by the U.S. Department of Energy Office of Science, Office of Nuclear Physics under Award Numbers DE-FG02-97ER41020 and DE-SC0019304. We further acknowledge support from the PRISMA Cluster of Excellence at the University of Mainz, the ‘Helmholtz Alliance for Astroparticle Physics HAP’ funded by the Initiative and Networking Fund of the Helmholtz Association, the Institute of Experimental Particle Physics at the Karlsruhe Institute of Technology (the Research University in the Helmholtz Association), and the German Academic Exchange Service.

\end{acknowledgments}

%\section{To Do}
%\medskip

\bibliography{TRIMS}
\bibliographystyle{apsrev}
\end{document}